\documentclass[10pt,a4paper, twoside]{article}
\usepackage[T1]{fontenc}
\usepackage{polski}
\usepackage[polish,british]{babel}
\usepackage[utf8]{inputenc}
\usepackage{graphicx}
\usepackage[usenames,dvipsnames,svgnames,table]{xcolor}
\usepackage{color}
\usepackage{float} 
\usepackage{amsmath}
\allowdisplaybreaks
\usepackage{amsfonts}
\usepackage{bm}
\usepackage{dsfont}
\usepackage{cancel}
\usepackage[margin=1in]{geometry}
\usepackage{caption}
\usepackage{subcaption}
\usepackage{lscape}
\usepackage{pdfpages} 
\usepackage{fancyhdr} 
\usepackage{slashed}  
\usepackage{csquotes} 
\date{}
\usepackage{lastpage}
\usepackage{booktabs}

\def\ltap{\raisebox{-.6ex}{\rlap{$\,\sim\,$}} \raisebox{.4ex}{$\,<\,$}} 
\def\gtap{\raisebox{-.6ex}{\rlap{$\,\sim\,$}} \raisebox{.4ex}{$\,>\,$}}

\makeatletter

 \renewcommand\@seccntformat[1]{\csname the#1\endcsname.\quad}

\makeatother

\usepackage{cite,./mcite}   

\begin{document}

\begin{titlepage}

\begin{flushright}
  CERN-TH-2019-130 
\end{flushright}

\par \vspace{10mm}

\parbox[t]{0.93\textwidth}{\centering\Large \bfseries  
Dynamical resolution scale    
 in 
  transverse momentum distributions at the LHC}
\vspace{10mm}

\begin{center}
{\bf F. Hautmann${}^{a, b, c, d},$ 
L. Keersmaekers${}^{a},$ 
A. Lelek${}^{a}$ 
and A. M. van Kampen${}^{a}$\\
}

\vspace{5mm}

$^{a}$ Elementaire Deeltjes Fysica, Universiteit Antwerpen, B 2020 Antwerpen
 
$^{b}$  Theoretical Physics 
Department, University of Oxford,  Oxford OX1 3NP

$^{c}$  CERN,  Theoretical Physics, CH-1211 Geneva 23

$^{d}$  UPV/EHU University of the Basque Country,  Bilbao E 48080 

\vspace{5mm}
\end{center}

\par \vspace{2mm}
\begin{center} {\bf Abstract} \end{center}
\begin{quote}
\pretolerance 10000

The QCD evolution of transverse momentum dependent (TMD) distribution functions has recently been formulated in a 
parton branching (PB) formalism. In this approach, soft-gluon coherence effects are taken into account by introducing 
the soft-gluon resolution scale and exploiting the  relation between transverse-momentum recoils and branching 
scales. In this work we investigate the implications of dynamical, i.e., branching scale dependent, resolution scales. 
We present both analytical studies and numerical solution of PB evolution equations in the presence of dynamical resolution scales. 
We use this to compare PB results with other approaches in the literature, and to    
analyze predictions for transverse momentum distributions in $Z$-boson production at the Large Hadron Collider (LHC).

\end{quote}

\end{titlepage}
 
\section{Introduction}

Theoretical predictions for precision physics at high-energy hadron colliders require methods for QCD resummations~\cite{Luisoni:2015xha}  
to all orders in the strong coupling. For observables sensitive to Sudakov resummation,  transverse momentum dependent (TMD) 
parton distribution and decay functions~\cite{Angeles-Martinez:2015sea} provide a theoretical framework   to both  
carry out resummed perturbative calculations and  incorporate systematically nonperturbative dynamics. 
 
In Refs.~\cite{Hautmann:2017xtx,Hautmann:2017fcj} a method has been proposed to treat TMDs in a parton branching (PB) formalism. 
The method is based on the 
unitarity picture of parton evolution~\cite{Webber:1986mc,R.K.Ellis2003}, and   takes into account color coherence of 
soft-gluon radiation~\cite{Bassetto:1984ik,Dokshitzer:1987nm,Marchesini:1987cf,Catani:1990rr} 
and transverse momentum recoils.   It introduces  the soft-gluon resolution scale  
to separate resolvable and non-resolvable branchings, and  Sudakov form factors  to express partonic  
probabilities for no resolvable branchings  in a given evolution interval.  

An important point in obtaining TMD distributions from the PB method concerns the ordering variables used to perform the branching evolution. 
 Because the transverse momentum generated radiatively in the branching  is sensitive to the 
treatment of the non-resolvable region~\cite{Hautmann:2007uw}, a supplementary condition  is needed to relate 
the transverse momentum recoil and the scale of the branching. This relation embodies the well-known property of 
angular ordering, and implies that the soft-gluon resolution scale can be dynamical,  i.e.,  dependent on the 
branching scale. 

In this paper we investigate the effects of dynamical resolution scales on TMD evolution and on collider observables. 
Using a mapping between branching scales and transverse momenta, we 
 discuss the resolvable radiation  regions and  PB evolution equations.  
 We solve these equations with dynamical resolution scale numerically by applying the   Monte Carlo solution techniques developed 
 in~\cite{Hautmann:2017xtx,Hautmann:2014uua}. 
We compare the PB results with results from two other approaches: the coherent branching approach 
of~\cite{Marchesini:1987cf,Catani:1990rr} (CMW) and the single-emission approach 
of~\cite{Kimber:1999xc,Kimber:2001sc,Watt:2003mx,Martin:2009ii}  (KMRW).  
 We present an application of our formalism to  
the  $Z$-boson transverse momentum distribution in  Drell-Yan (DY)  production~\cite{Drell:1970wh} at the LHC, and study its sensitivity to dynamical resolution scales 
at low transverse momenta.

The paper is organized as follows. In Sec.~2 we recall the basic elements of the PB approach,  introduce the dynamical 
soft-gluon resolution scale, and describe the resolvable and non-resolvable emission regions. In Sec.~3 we map   
branching scales  to transverse momenta, and give  the corresponding form of PB  equations.  
In Sec.~4 we use these results to perform analytic comparisons of multiple-emission and single-emission TMD approaches. 
We compare PB results with KMRW and CMW results. 
In Sec.~5 we solve the PB evolution equation with dynamical resolution scale by numerical     
methods, and  present predictions for  the $Z$-boson 
transverse momentum spectrum at the LHC.  We give conclusions in Sec.~6.

\section{PB TMDs and soft-gluon angular ordering}

In this section we summarize the main elements of TMD evolution in the PB formalism, stressing  in particular the aspects 
associated with soft-gluon  angular ordering.  

In the  PB  approach~\cite{Hautmann:2017xtx,Hautmann:2017fcj}  the TMD evolution equations can be written  as
\begin{eqnarray}
&& \widetilde{A}_a\left( x, {\bm k}, \mu^2\right) = \Delta_a\left(\mu^2, \mu_0^{2}\right)\widetilde{A}_a\left( x, {\bm k}, \mu_0^2\right)+ 
 \sum_b\int \frac{\textrm{d}^2{\boldsymbol \mu}^{\prime}}{\pi {\mu}^{\prime 2}}\Theta\left(\mu^{2}-\mu^{\prime 2}\right)\Theta\left(\mu^{\prime 2}-\mu_0^{ 2}\right)
 \nonumber \\ 
 &\times& 
  \int_x^{1}\textrm{d}z \ \Theta\left(z_M (\mu^\prime) - z \right)
  {  { \Delta_a\left(\mu^2, \mu_0^2  \right)  } \over 
  { \Delta_a\left(\mu^{\prime 2}, \mu_0^2 \right) } } \ 
  P_{ab}^{R}\left(z,\alpha_s(b(z)^2\mu^{\prime 2})\right)\widetilde{A}_b\left( \frac{x}{z},  {\bm k} + a(z){\boldsymbol \mu}^\prime, \mu^{\prime 2}\right) \; , 
\label{eq:tmdevol}
\end{eqnarray} 
where $\widetilde{A}_a\left( x, {\bm k}, \mu^2\right)= x A_a\left( x, {\bm k}, \mu^2\right)$ is the momentum-weighted 
TMD distribution of flavor $a$, carrying the longitudinal momentum  fraction $x$ of the hadron's momentum and  transverse momentum ${\bm k}$\footnote{We use 
the   notation $k=(k^0, k^1, k^2, k^3)=(E_{k}, {\bm k}, k^3)$, where ${\bm k}=(k^1, k^2)$, and $k_{\bot} = | {\bm k} | $. } 
at the evolution scale $\mu$;  $z$ and ${\boldsymbol \mu}^\prime$ are the branching variables, with $z$ being the longitudinal momentum transfer  
at the  branching, and  $ \mu^\prime = \sqrt{ {\boldsymbol \mu}^{\prime 2}}$ the momentum scale at which the branching occurs; 
 $P_{ab}^R$ are the real-emission splitting kernels; 
 $\Delta_a$ is   the Sudakov form factor,  given by  
\begin{equation}
\label{sudakovexpression}
\Delta_a(\mu^2, \mu_0^2)  = \exp\left[-\sum_b \int_{\mu_0^2}^{\mu^2}\frac{\textrm{d}\mu^{\prime 2}}{\mu^{\prime 2}} \int_0^{1}\textrm{d}z \ \Theta\left(z_M (\mu^\prime) - z \right) 
\ z \ P_{ba}^{R}\left(z,\alpha_s\left(b(z)^2\mu^{\prime 2}\right)\right) \right]\;.
\end{equation}
  The initial  
evolution scale is denoted by $\mu_0$. 

The functions $a(z)$, $b(z)$ and 
$z_M (\mu^\prime)$ in 
Eqs.~(\ref{eq:tmdevol})  and  (\ref{sudakovexpression})  
encode features  associated with the  ordering variables used to perform the branching evolution, and  are  specified below. 

\begin{figure}[htbp]
\centering 
\includegraphics[width=0.5\textwidth]{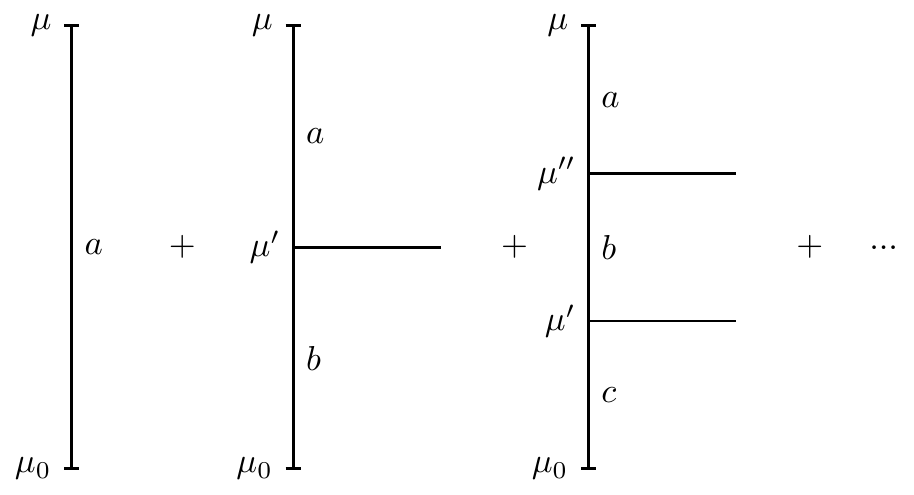}
\caption{Solution of the branching equation by iteration.}  
 \label{Fig:evolution}
\end{figure}

An iterative Monte Carlo solution of Eq.~(\ref{eq:tmdevol}) is obtained 
in Ref.~\cite{Hautmann:2017fcj}, and is represented pictorially in Fig.~\ref{Fig:evolution}.  The distribution of flavor $a$ at scale $\mu$ is written,  
as a function of $x$ and ${\bm k}$,  
 as a sum of terms involving, iteratively,  no branching  between $\mu_0$ and $\mu$, then one branching, then two branchings, and so forth.  
 The transverse momentum ${\bm k}$, in particular, arises from this solution by 
 combining the intrinsic transverse momentum (in the first term on the right hand side  of Eq.~(\ref{eq:tmdevol})) with  the transverse momenta 
 emitted at all branchings.

Let us  examine    $a(z)$, $b(z)$ and 
$z_M (\mu^\prime)$ in 
Eqs.~(\ref{eq:tmdevol})  and  (\ref{sudakovexpression}).    
The function $a(z)$ in the last factor on the right hand side of  Eq.~(\ref{eq:tmdevol})  gives the relation between the scale  
  of the branching and the transverse momentum 
    of the emitted parton.   In the case of angular ordering one has~\cite{Hautmann:2017fcj}  
\begin{equation} 
\label{a-function-ao} 
 a(z) = 1 - z \;\; , 
\end{equation}     
and the transverse momentum $q_\perp$ 
    of the emitted parton is related to the branching scale $ \mu^{\prime} $   by~\cite{Hautmann:2017fcj}  
 \begin{equation}
 \label{eq:qtandmuAO}
 q_{\bot}^2 = (1-z)^2\mu^{\prime 2}\;.
 \end{equation}     

The function $b(z)$ 
specifies  the momentum scale 
in the running coupling $\alpha_s$. For angular ordering one has~\cite{Marchesini:1987cf,Catani:1990rr,Martinez:2018jxt}   
\begin{equation} 
\label{b-function-ao} 
 b(z) = 1 - z \;\; , 
\end{equation} 
so that the coupling is evaluated at the transverse momentum scale, 
\begin{equation} 
\label{alphas-ao} 
\alpha_s\left(b(z)^2\mu^{\prime 2}\right) =  \alpha_s   \left( q_\perp^2 \right)   \;\; .  
\end{equation}   

The function  $z_M (\mu^\prime)$  specifies   the 
   soft-gluon resolution scale~\cite{Hautmann:2017xtx} which separates  
the region of   resolvable branchings ($z < z_M$) from the region of  
  non-resolvable  branchings ($z > z_M$), for any given $\mu^\prime$. 
  Let us denote by $q_0$ the  minimum 
transverse momentum 
with which    any  emitted parton    can be resolved, so that 
\begin{equation} 
\label{minqT} 
q_\perp > q_0 \; .  
\end{equation} 
 By inserting the angular ordering relation  (\ref{eq:qtandmuAO}) 
 into Eq.~(\ref{minqT}),   the 
 condition for  resolving soft gluons  is given by $z < z_M (\mu^\prime) $ with~\cite{Webber:1986mc,R.K.Ellis2003,Hautmann:2017fcj}  
\begin{equation}
\label{eq:zmAO}
z_M  (\mu^\prime) = 1-  {q_0} / {\mu^{\prime }} \; ,   
\end{equation}
 where the momentum scale $q_0$ is understood   to be 
 $q_0 \gtap \Lambda_{\rm{QCD}}$. 
  
The role of the functions $a(z)$ in Eq.~(\ref{a-function-ao}) and    $b(z)$  in Eq.~(\ref{b-function-ao})  has 
 been analyzed   in 
 Refs.~\cite{Martinez:2018jxt,Martinez:2019mwt,Hautmann:2019rvr} 
 for TMD applications  to  DY  processes.  
  In this work we concentrate  on    implications  of  the resolution scale  $z_M (\mu^\prime)$  in Eq.~(\ref{eq:zmAO}).

By integrating the TMD distributions in Eq.~(\ref{eq:tmdevol}) over transverse momenta one obtains collinear initial-state distributions, 
\begin{equation} 
 \widetilde{f}_a\left(x, \mu^2\right) =  \int { {\textrm{d}^2{\bm k}  } \over \pi } \ \widetilde{A}_a\left(x, {\bm k}, \mu^2\right) \;.
\label{integratingAoverKt}
\end{equation}
 It has been shown 
in~\cite{Hautmann:2017xtx,Hautmann:2017fcj} that for  
$z_M \to 1$ and $\alpha_s \to \alpha_s (\mu^{\prime 2})$  these are   collinear parton distribution functions satisfying  
DGLAP evolution equations~\cite{Gribov:1972ri,Lipatov:1974qm,Altarelli:1977zs,Dokshitzer:1977sg}.\footnote{The convergence to DGLAP 
 at leading order (LO) and  next-to-leading order (NLO)  has been 
verified numerically in~\cite{Hautmann:2017fcj} against the evolution program~\cite{Botje:2010ay} 
  at  level of better   than 1\% over a range of five orders of magnitude both in $x$ and in $\mu$.}   On the other hand,     
for general $z_M$ and $\alpha_s$  of the form in  Eq.~(\ref{eq:zmAO}) and  Eq.~(\ref{alphas-ao})     the evolution equation for $ \widetilde{f}_a $ 
is given  by 
\begin{eqnarray}
\label{PBangular}
  &&\widetilde{f}_{a}(x,\mu ^{2}) = 
  \Delta_a(\mu ^{2} , \mu_0^2)    
  \widetilde{f}_{a}(x,\mu _{0} ^{2})  + 
  \sum _{b}  \int_{\mu_{0} ^{2}}^{ \mu ^{2}}  \frac{d \mu ^{\prime 2}}{\mu ^{\prime 2}}     \int _{x}^{1}  dz  
   \nonumber \\ &\times& 
   \Theta ( 1  - 
 q_0 / \mu^\prime - z )   \ 
  \frac{\Delta_a(\mu ^{2}, \mu_0^2)}{\Delta_a(\mu ^{\prime 2}, \mu_0^2)} \  
  P^{R}_{ab}\left(z, \alpha_s\left((1-z)^2\mu^{\prime 2}\right)\right)\widetilde{f}_{b}\left(\frac{x}{z},\mu ^{\prime 2}\right)  \;\;.
\end{eqnarray} 

The kernels of the evolution equations (\ref{eq:tmdevol}) and (\ref{PBangular}) have support in the resolvable emission 
region $ x < z < z_M$. We depict  this region in  the $(\mu^\prime , z)$ plane  in Fig.~\ref{fig:integralLimitszmu},  with the 
dynamical  resolution scale (\ref{eq:zmAO}).  Fig.~\ref{fig:integralLimitszmu}(a) represents the case of contributions 
to the distribution function with  $x \geq 1- q_0 / \mu_0$, while Fig.~\ref{fig:integralLimitszmu}(b) 
represents the case of  $ x < 1- q_0 / \mu_0$.

\section{Mapping evolution scales to transverse momenta}

\begin{figure}[!htb]
    \centering
        \includegraphics[width=16cm]{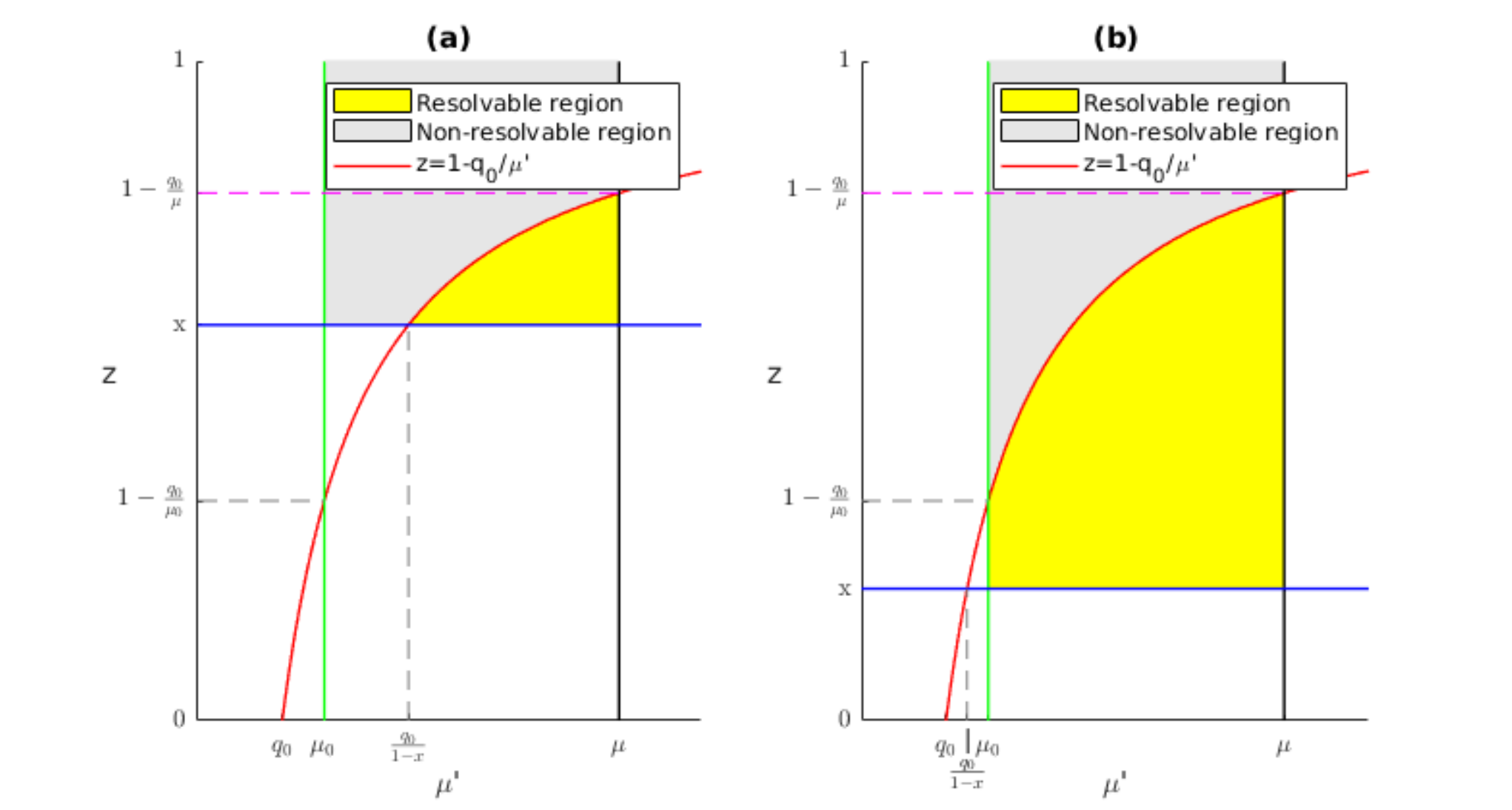}
        \caption{The angular ordering condition $z_M (\mu^\prime) = 1 - q_0 / \mu^\prime$  with the resolvable and non-resolvable 
emission regions in  the $(\mu^\prime , z)$ plane: 
a) the case  $1 > x \geq 1- q_0 / \mu_0$; b) the case  $   1- q_0 / \mu_0 > x > 0 $.}
     \label{fig:integralLimitszmu}
\end{figure}

We next recast the parton-branching evolution and 
separation between resolvable and non-resolvable 
branchings in terms of longitudinal momentum fractions and 
transverse momenta. To this end, we exploit the angular ordering relation in Eq.~(\ref{eq:qtandmuAO}) to map  
branching  scales  on to transverse momenta 
for the resolvable regions  
in Fig.~\ref{fig:integralLimitszmu}. 

Given the minimum transverse momentum 
$q_0$ and the lowest  scale $\mu_0$ of the branching, 
for any $x$ it is useful to distinguish the two cases 
illustrated in  Fig.~\ref{fig:integralLimitszmu}, 
depending on whether   
a) $\mu_0 \leq q_0 / (1 - x) $ or 
b) $\mu_0 > q_0 / (1 - x) $. 
For any $z$ with $ x \leq z \leq 1$,   
in case a) the emitted transverse momentum 
spans the interval 
$ q_0  \leq q_\perp \leq 
 \mu (1 - z)$, while in case b) we have 
$ \mu_0 (1 - z) \leq q_\perp \leq 
 \mu (1 - z)$. 
This results into different forms of the branching equations in the two cases, once they are expressed 
directly in terms of transverse momenta. 

\subsection{Case a) $1 > x \geq 1-{q_0} / {\mu_0}$}

For $x  \geq    1-{q_0}  /  {\mu_0}$ the resolvable 
emission region is mapped to the domain     in the $(z , q_{\bot})$ plane  
pictured in Fig.~\ref{fig:integralLimitszqt}(a).  
We  change 
integration variable from $\mu^\prime $ to $q_\perp$ 
in the  branching equation  (\ref{PBangular})  using 
 the angular ordering relation   (\ref{eq:qtandmuAO}). Then 
 Eq.~(\ref{PBangular}) 
can be recast in terms of transverse momenta as 
\begin{eqnarray}
\label{PBangular_2term_3}
 \widetilde{f}_{a}(x,\mu ^{2}) &=& 
 \Delta_a(\mu ^{2} , \mu _{0} ^{2}) 
 \widetilde{f}_{a}(x,\mu _{0} ^{2})+  
  \sum _{b}  \int   \frac{\textrm{d}q_{\bot}^2}{q_{\bot}^2}
  \int_{x}^{1}\textrm{d}z \ 
 \Theta ( q_\perp^2 - q_0^2 ) \ \Theta ( \mu^2 
(1-x)^2  - q_\perp^2)   
  \nonumber \\ 
&\times&  \Theta ( 1  - q_\perp / \mu -z ) \ 
 \frac{\Delta_a(\mu ^{2}, \mu_0^2)}{\Delta_a\left(q_{\bot}^2 / (1-z)^2,\mu_0^2\right)}  \ 
  P^{R}_{ab}\left(z, \alpha_s\left(q_{\bot}^2\right)\right)\widetilde{f}_{b}\left(\frac{x}{z},\frac{q_{\bot} ^{2}}{(1-z)^2}\right) \; .  
\end{eqnarray}

\subsection{ 
Case b)  $  1- {q_0} / {\mu_0} > x > 0 $}    

\begin{figure}[!htb]
        \centering
        \includegraphics[width=16cm]{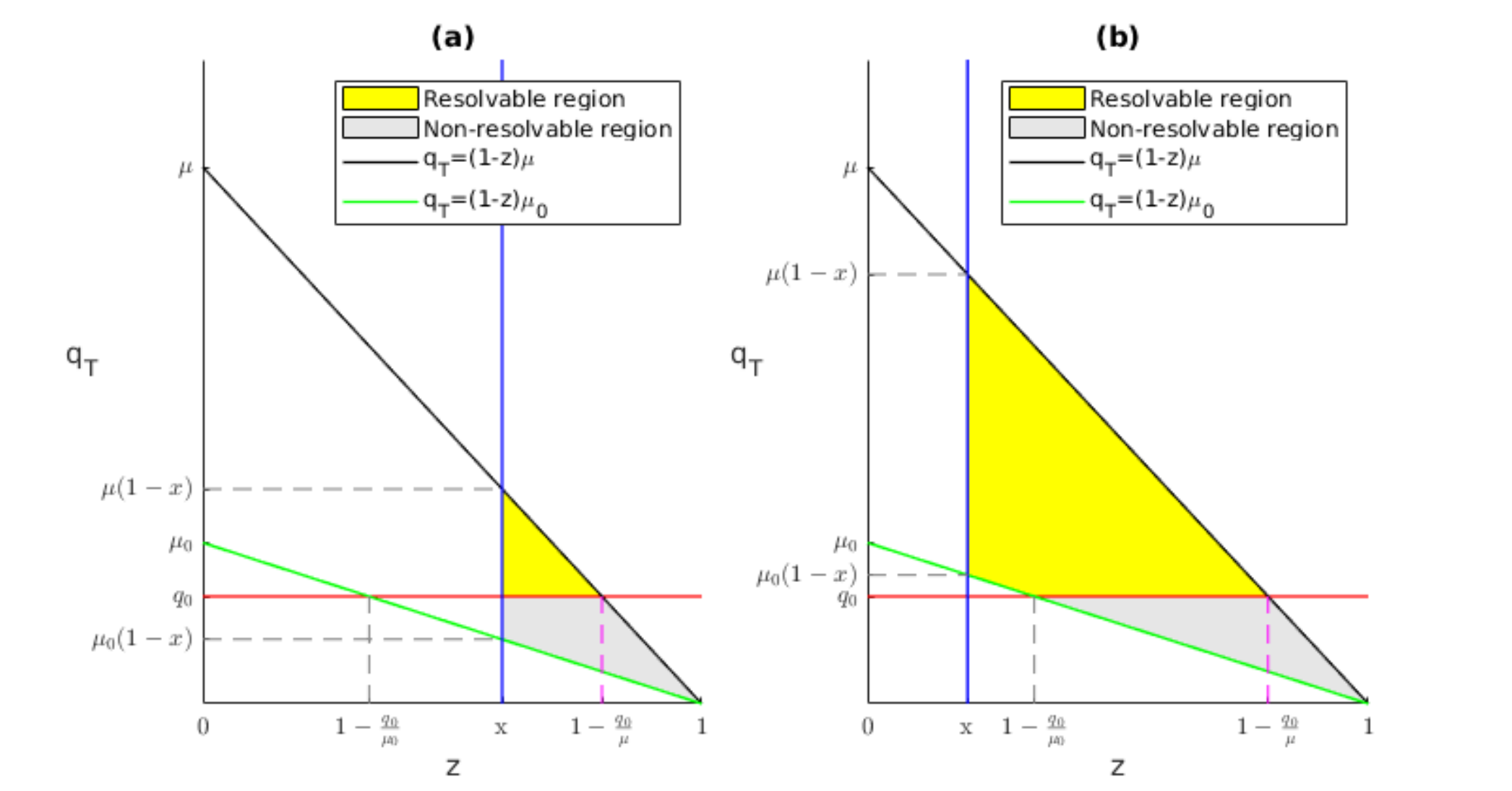}
        \caption{Resolvable and non-resolvable emission regions in the $(z , q_{\bot})$ plane for evolution in the 
cases   a)   $1 > x \geq 1- q_0 / \mu_0$ and b)  $   1- q_0 / \mu_0 > x > 0 $. }
        \label{fig:integralLimitszqt}
\end{figure}

For $x < 1-{q_0}  /  {\mu_0}$ the resolvable 
emission region is mapped to the domain     in the $(z , q_{\bot})$ plane  
pictured in Fig.~\ref{fig:integralLimitszqt}(b).  
Performing the same change of integration variable in  
Eq.~(\ref{PBangular})  as in case a) of the previous subsection,  
we recognize that  now a subtraction term arises from the 
low-$q_\perp$  region,  $q_0 <  q_\perp < (1-x) \mu_0 $, so that  
Eq.~(\ref{PBangular}) 
is rewritten  in terms of transverse momenta as 
\begin{eqnarray}
\label{eq:aofinalcase2}
 \widetilde{f}_{a}(x,\mu ^{2}) &=& 
 \Delta_a(\mu ^{2} , \mu _{0} ^{2}) 
 \widetilde{f}_{a}(x,\mu _{0} ^{2})+  
  \sum _{b}  \int   \frac{\textrm{d}q_{\bot}^2}{q_{\bot}^2}
  \int_{x}^{1}\textrm{d}z \  \Big[ 
 \Theta ( q_\perp^2 - q_0^2 ) \ \Theta ( \mu^2 
(1-x)^2  - q_\perp^2)    
  \nonumber \\ 
&\times&    \Theta ( 1  - q_\perp / \mu -z  ) -  \Theta ( q_\perp^2 -  q_0^2 ) \ \Theta ( \mu_0^2 
(1-x)^2  - q_\perp^2) \  
 \Theta ( 1  - q_\perp / \mu_0 -z )     \Big]       
  \nonumber \\ 
&\times&
 \frac{\Delta_a(\mu ^{2}, \mu_0^2)}{\Delta_a\left(q_{\bot}^2 / (1-z)^2,\mu_0^2\right)}  \ 
  P^{R}_{ab}\left(z, \alpha_s\left(q_{\bot}^2\right)\right)\widetilde{f}_{b}\left(\frac{x}{z},\frac{q_{\bot} ^{2}}{(1-z)^2}\right) \; .  
\end{eqnarray}
 We observe that the first term in the 
square bracket in Eq.~(\ref{eq:aofinalcase2})  is a contribution   analogous to that in 
Eq.~(\ref{PBangular_2term_3}), while    the second term in the 
square bracket  provides the  low-$q_\perp$  subtraction.

Alternatively, the branching kernel in the case $x < 1-{q_0}  /  {\mu_0}$  can be expressed as a sum of two contributions,  
corresponding  respectively to the $q_\perp < (1-x) \mu_0 $ region and 
$q_\perp > (1-x) \mu_0 $ region, as follows   
\begin{eqnarray}
\label{case-a}
 \widetilde{f}_{a}(x,\mu ^{2}) &=& 
 \Delta_a(\mu ^{2} , \mu _{0} ^{2}) 
 \widetilde{f}_{a}(x,\mu _{0} ^{2})+  
  \sum _{b}  \int   \frac{\textrm{d}q_{\bot}^2}{q_{\bot}^2}
  \int_{x}^{1}\textrm{d}z    \Big[  
 \Theta ( q_\perp^2 - q_0^2 ) \Theta ( \mu_0^2 
(1-x)^2  - q_\perp^2)    
 \nonumber \\ 
&\times& 
 \Theta ( z + q_\perp / \mu_0 -1 ) 
 \Theta ( 1  - q_\perp / \mu -z ) 
+   \Theta ( q_\perp^2 -  (1-x)^2 \mu_0^2 ) \Theta ( \mu^2 
(1-x)^2  - q_\perp^2) 
  \nonumber \\ 
&\times&    \Theta ( 1  - q_\perp / \mu  -z )   
 \Big] 
 \frac{\Delta_a(\mu ^{2}, \mu_0^2)}{\Delta_a\left(q_{\bot}^2 / (1-z)^2,\mu_0^2\right)}   
 P^{R}_{ab}\left(z, \alpha_s\left(q_{\bot}^2\right)\right)\widetilde{f}_{b}\left(\frac{x}{z},\frac{q_{\bot} ^{2}}{(1-z)^2}\right) \; . 
\end{eqnarray}
Here the two terms of the sum in the square bracket, given by  products of $\Theta$ functions,   
describe the  low-$q_\perp$  and  high-$q_\perp$  contributions.

In the next section we will use the form for  the branching equations derived above,   along with the formulas of 
Sec.~2, 
to carry out  a comparison of the 
PB method~\cite{Hautmann:2017xtx,Hautmann:2017fcj} 
with other 
existing approaches in the literature.   We will  
analyze in particular different treatments of the QCD parton  cascade,  in which the transverse momentum is 
  generated either through  multiple emissions or through  a single emission.

\section{Multiple-emission versus single-emission 
approaches}

As an approach  based on the 
unitarity picture~\cite{Webber:1986mc,R.K.Ellis2003} of 
parton evolution and angular ordering, the PB 
method~\cite{Hautmann:2017xtx,Hautmann:2017fcj} can naturally be compared with the coherent branching 
approach~\cite{Marchesini:1987cf,Catani:1990rr}  of 
Catani-Marchesini-Webber (CMW). 
Since Refs.~\cite{Marchesini:1987cf,Catani:1990rr} do 
not  construct  TMD distributions, we   examine  
branching equations in the 
PB  and CMW approaches at the level of integrated distributions.  At this level, we 
 observe that Eq.~(\ref{PBangular}) agrees with the CMW 
result --- see Eqs.~(42), (49)  
and  Sec.~3.4  of~\cite{Marchesini:1987cf}.   In Ref.~\cite{Marchesini:1987cf}  this  branching equation is studied at LO with one-loop splitting  
kernels and running coupling, while in  
Ref.~\cite{Hautmann:2017xtx} (and in the present paper) it 
is   studied  at NLO with two-loop splitting kernels and running coupling.\footnote{The treatment in 
Ref.~\cite{Hautmann:2017xtx} 
 incorporates in particular the  two-loop correction  
to the coupling which is shown in~\cite{Catani:1990rr} to be required to obtain next-to-leading-logarithmic accuracy in the soft-gluon resummation.}   

The Kimber-Martin-Ryskin-Watt (KMRW) approach \cite{Kimber:1999xc,Kimber:2001sc,Watt:2003mx,Martin:2009ii}, on the other hand, is designed to 
construct TMD unintegrated distributions. 
In contrast to the PB method, in which  the transverse momentum and the branching scale are   calculated at each branching  as  illustrated   
in Fig.~\ref{Fig:evolution}, 
KMRW is a one-step evolution approach: it performs  evolution in one scale up to $q_{\bot}^2$,  while  
 the second scale is generated only in the last step of the evolution. The KMRW   physical picture is thus quite different from 
 that of PB and CMW. In particular, in KMRW the transverse momentum is produced as a result of a single emission, while in PB it is built from 
 multiple emissions.  
 
 We next perform comparisons of KMRW and PB 
 distributions. In the KMRW literature, 
  the distinction  between the values of the two momentum scales $\mu_0$ and $q_0$ discussed in Sec.~3 is not made. For the purpose of this comparison, therefore, 
  we set $ q_0 \approx \mu_0$ in the formulation of Sec.~3, and we will thus be using the branching equation valid in case a) of 
  Subsec.~3.1,  Eq.~(\ref{PBangular_2term_3}).

\subsection{Comparison with the KMRW approach}

In the KMRW approach  the TMD distribution is written 
as~\cite{Kimber:1999xc,Kimber:2001sc,Watt:2003mx,Martin:2009ii}
\begin{eqnarray}
\label{eq:KMRWTMD}
\widetilde{D}_a(x,\mu ^{2}, q_{\bot}^2)&=&T_a(\mu ^{2}, q_{\bot} ^{ 2})\sum _{b} \int _{x}^{1-C(q_{\bot}, \mu)} dz P^{R}_{ab}\left(z, \alpha_s(q_{\bot} ^{ 2})\right)\widetilde{f}_{b}\left(\frac{x}{z}, q_{\bot}^{ 2}\right) \;\;  , 
\end{eqnarray}
where  the Sudakov form factor is given by 
\begin{eqnarray}
\label{eq:sudakovkmrw}
T_a(\mu^2, q_{\bot}^2) = \exp\left[-\int_{q_{\bot}^2}^{\mu^2}\frac{\textrm{d}q_{\bot}^{\prime 2}}{q_{\bot}^{\prime 2}}\sum_b \int_0^{1-C(q_{\bot}^{\prime}, \mu)}\textrm{d}z z P_{ba}^{R}\left(z,\alpha_s\left(q_{\bot}^{\prime 2}\right)\right) \right]\; , 
\end{eqnarray}
and the collinear density   $\widetilde{f}_{a}(x,\mu ^{2}) $ 
obeys the evolution equation 
\begin{eqnarray}
\label{eq:KMR}
  \widetilde{f}_{a}(x,\mu ^{2}) &=& \widetilde{f}_{a}(x,\mu _{0} ^{2})T_a(\mu ^{2}, \mu_0^2)    \nonumber \\ &+& \int_{\mu_{0} ^{2}}^{q_{\bot M}^2}  \frac{d q_{\bot} ^{ 2}}{q_{\bot} ^{ 2}}\left(T_a(\mu ^{2}, q_{\bot} ^{ 2})\sum _{b} \int _{x}^{1-C(q_{\bot}, \mu)} dz P^{R}_{ab}\left(z, \alpha_s(q_{\bot} ^{ 2})\right)\widetilde{f}_{b}\left(\frac{x}{z}, q_{\bot}^{ 2}\right) \right)   \; . 
\end{eqnarray}
The phase space parameters $ C(q_{\bot}, \mu)$ and  $ q_{\bot M}$ in the above formulas are assigned 
according to two distinct prescriptions~\cite{Kimber:1999xc,Kimber:2001sc,Watt:2003mx,Martin:2009ii,Golec-Biernat:2018hqo} in the KMRW approach:  
\begin{equation}
\label{kmrw-ps-so} 
 C(q_{\bot}, \mu) = {q_{\bot}}  /  {\mu} \;\; , \;\;\;\;  q_{\bot M}  = \mu(1-x) \;\;\;\;   {\rm{for}} \;\; {\rm{KMRW}} \;\;  {\rm{strong}} \;\; {\rm{ordering}}  
\end{equation}
and 
\begin{equation} 
\label{kmrw-ps-ao}
C(q_{\bot}, \mu) = {q_{\bot}} /  ({q_{\bot}+ \mu})   \;\; , \;\;\;\;  q_{\bot M}  =  \mu({1-x})  / {x}  \;\;\;  {\rm{for}}  \;\; {\rm{KMRW}} \;\;  {\rm{angular}} \;\; {\rm{ordering}}   . 
\end{equation}

Having mapped the PB evolution onto 
transverse momenta in Sec.~3, we are in 
a position to  directly compare the PB and 
KMRW results.    By considering 
Eq.~(\ref{PBangular_2term_3}) and Eq.~(\ref{eq:KMR}) with KMRW strong ordering conditions (\ref{kmrw-ps-so}),  
we recognize that  PB and KMRW 
differ in the momentum scales  at which  both the Sudakov form factor and the collinear density $\widetilde{f}_b$ are 
evaluated, as  KMRW uses  transverse momenta whereas PB uses transverse momenta rescaled by 
$1/(1-z)$. From  
Eq.~(\ref{PBangular_2term_3}) and Eq.~(\ref{eq:KMR}) with KMRW angular ordering conditions (\ref{kmrw-ps-ao}),  
we recognize that in this case 
PB and KMRW, besides differing in the 
arguments of Sudakov factor and collinear density, differ 
also in the phase space regions 
in longitudinal and transverse momenta that are populated 
by the radiative processes. 

We thus see that, also taking into account  
 the possible prescriptions in Eqs.~(\ref{kmrw-ps-so}) and (\ref{kmrw-ps-ao}), 
the one-step picture of KMRW leads to different results  from 
the multiple-emission PB picture.  
In Sec.~5 we illustrate the implications of these   
differences   
by performing numerical calculations for the 
TMD distributions that result from evolution in the 
two approaches, and examining the corresponding predictions 
for the DY $Z$-boson  transverse momentum 
spectra at the LHC.

\subsection{Remark on Sudakov form factors}

It is worth noting that the definition of the Sudakov form 
factor itself plays a different role in the context 
of the PB approach and the KMRW approach. 

In the PB approach 
the Sudakov form factor 
$ \Delta_{a}(\mu ^{2},\mu_0^2)$ 
has the interpretation of 
 probability for no resolvable branching in a given 
evolution interval from $\mu_0$ to $\mu$, and fulfills the 
property  
\begin{equation}
\label{sudsud}
 \Delta_{a}(\mu ^{2}, \widetilde{\mu} ^{2})\Delta_{a}(\widetilde{\mu} ^{2}, \mu_0^2) = \Delta_{a}(\mu ^{2},\mu_0^2)  \; 
\end{equation}
for any evolution scale $\widetilde{\mu}$. 

For example, 
for the  Sudakov form factor in the angular-ordered  
evolution we use  
\begin{equation}
\Delta_{a}(\mu^{2}  ,  \mu_0^{2})=\exp \left(- \sum_{b} 
\int _{\mu _{0}^{2}}^{\mu ^{2}} \frac{\textrm{d} \mu ^{\prime 2}}{\mu ^{\prime 2}}\int _{0}^{1-{q_0} / {\mu^{\prime}}} \textrm{d}z \ z \ P^{R}_{ba}\left(z, \alpha_s\left((1-z)^2\mu ^{\prime 2}\right)\right)\right)\;, 
\end{equation}
for which  Eq.~(\ref{sudsud}) is  fulfilled.
Upon mapping to transverse momenta, this  becomes
\begin{equation}
\Delta_a(\mu^2,\mu_0^2) = \exp\left[ - \sum_b  \left( 
\int_{\mu_0^2}^{\mu^2}\frac{\textrm{d}q_{\bot}^2}{q_{\bot}^2}\int_0^{1-{q_{\bot}} /  {\mu}}\textrm{d}z + 
\int_{q_0^2}^{\mu_0^2}\frac{\textrm{d}q_{\bot}^2}{q_{\bot}^2}\int_{1-{q_{\bot}} / {\mu_0}}^{1-{q_{\bot}} / {\mu}}\textrm{d}z
\right)  z \ P_{ba}^{R}\left(z,\alpha_s\left(q_{\bot}^{2}\right)\right)\right]\;, 
\end{equation}
for which  Eq.~(\ref{sudsud}) is still fulfilled.

On the other hand, using the   KMRW  expression in  Eq.~(\ref{eq:sudakovkmrw}),   Eq.~(\ref{sudsud}) is not 
fulfilled. Rather, one has 
\begin{equation}
T_{a}(\mu^{2}, k_{\bot}^2)T_{a}( k_{\bot}^2, \mu_0^2) = 
 T_a(\mu^2, \mu_0^2)\exp\left( \sum_b \int_{\mu_0^2}^{k_{\bot}^2}\frac{\textrm{d}q_{\bot}^2}{q_{\bot}^2}\int_{1-C(q_{\bot}, k_{\bot})}^{1-C(q_{\bot}, \mu)} \textrm{d}z \ 
 z \ P_{ba}^{R}(z, \alpha_s( q_{\bot}^2))\right)\;.
\end{equation}
This implies that, besides the different treatment of 
radiative processes noted in Subsec.~4.1,  we 
observe differences between the 
single-emission and multiple-emission approaches 
also in the treatment of 
the non-resolvable processes. We  may 
expect that the features noted in   
Subsec.~4.1 and in this subsection will lead to 
different behaviors 
in transverse momentum 
distributions  both at high transverse momenta and at  low 
transverse momenta.  

\section{Numerical results}

We  investigate next  the numerical implications of the 
analysis in the previous sections on TMD distributions 
and DY spectra. 

\subsection{TMDs from PB and KMRW}

 In this section we present numerical results for TMD distribution functions from the  
   PB approach with dynamical resolution scale.   
   We  perform  numerical comparisons with  KMRW TMDs. 
   The results are shown 
   as functions of flavor, longitudinal momentum fraction $x$, transverse momentum $k_\perp$, evolution 
 scale  $\mu$.

  KMRW TMD distribution sets have been obtained 
  in~\cite{Bury:2017jxo} according to the KMRW angular ordering prescription  (\ref{kmrw-ps-ao}), 
  using the CT10nlo PDF~\cite{Lai:2010vv} set as a 
   starting collinear distribution and a flat parameterization for $ k_{\bot} < 1\;\textrm{GeV}$ as an  intrinsic   $k_{\bot}$ distribution 
   at starting scale $\mu_0$.     
   These  distributions have been included in the TMDlib 
   library~\cite{Hautmann:2014kza} under 
   the name MRW-CT10nlo.\footnote{Strictly speaking, the TMD set MRW-CT10nlo has been obtained using the differential 
   definition of KMRW TMDs (see e.g. \cite{Bury:2017jxo, Golec-Biernat:2018hqo}). We have 
   performed also studies with KMRW TMDs defined according to the integral definition (as in Eq.~(\ref{eq:KMRWTMD})) and we have verified that 
   our conclusions remain valid.} 
   
   To evaluate  PB TMDs,  
   we solve numerically 
   Eq.~(\ref{eq:tmdevol}),   with the 
   angular ordering condition in Eq.~(\ref{eq:qtandmuAO}),  the 
    dynamical resolution scale in Eq.~(\ref{eq:zmAO}) where  we take 
    $q_0=1\;\textrm{GeV}$, and 
the     scale of $\alpha_s$  in Eq.~(\ref{alphas-ao}). 
    We    use   the  Monte Carlo  solution method 
  developed 
  in~\cite{Hautmann:2017xtx,Hautmann:2017fcj} and 
  implemented in the package  
  uPDFevolv~\cite{Hautmann:2014uua}.   
  Following~\cite{Martinez:2018jxt}, we take intrinsic  $k_{\bot}$ distribution given by a 
  simple  gaussian at starting scale $\mu_0$ 
  with (flavor-independent and $x$-independent)  width $\sigma= k_0 / \sqrt{2}$, $k_0=0.5$ GeV. 
   For the purpose of performing  comparisons with the KMRW TMD 
   set MRW-CT10nlo,   we 
    take the same starting collinear distribution 
    CT10nlo~\cite{Lai:2010vv}.   We also  note that  
  the infrared region with  the Landau pole of 
   the coupling in Eq.~(\ref{alphas-ao}) is avoided by using 
  the   dynamical resolution scale (\ref{eq:zmAO})  with   $q_0=1\;\textrm{GeV}$.

In addition, we introduce an approximation to the PB framework, which we refer to as ``PB-last-step'', which is obtained from PB by taking the same settings as the full PB calculation but 
restricting the transverse momentum  $k_{\bot}$  to the last emission only. We use the PB-last-step Monte Carlo simulation as a guidance to distinguish  effects from 
 single emission and multiple emissions.

\begin{figure}
\begin{minipage}{0.49\linewidth}
\centerline{\includegraphics[width=0.99\linewidth]{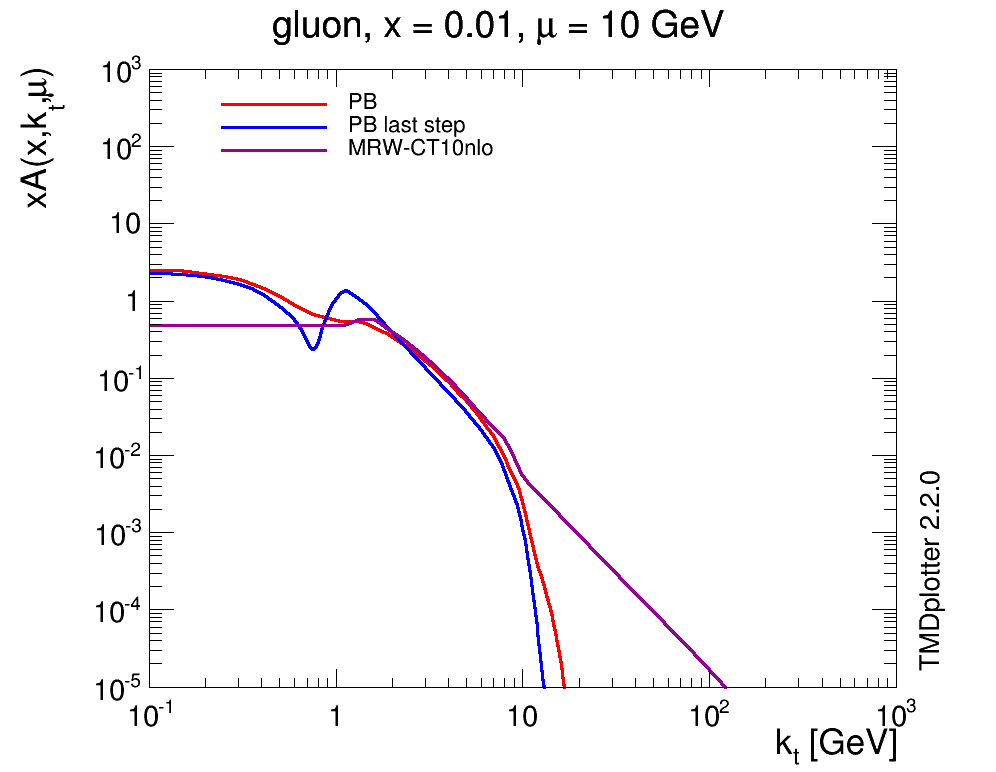}}
\end{minipage}
\hfill
\begin{minipage}{0.49\linewidth}
\centerline{\includegraphics[width=0.99\linewidth]{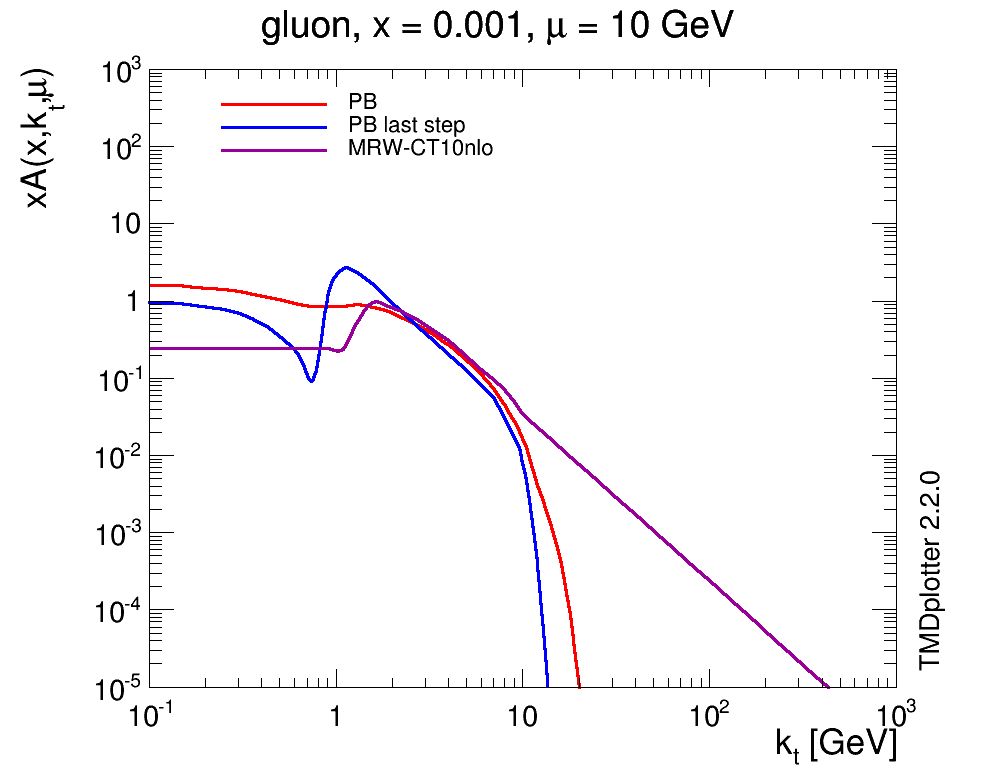}}
\end{minipage}
\hfill
\begin{minipage}{0.49\linewidth}
\centerline{\includegraphics[width=0.99\linewidth]{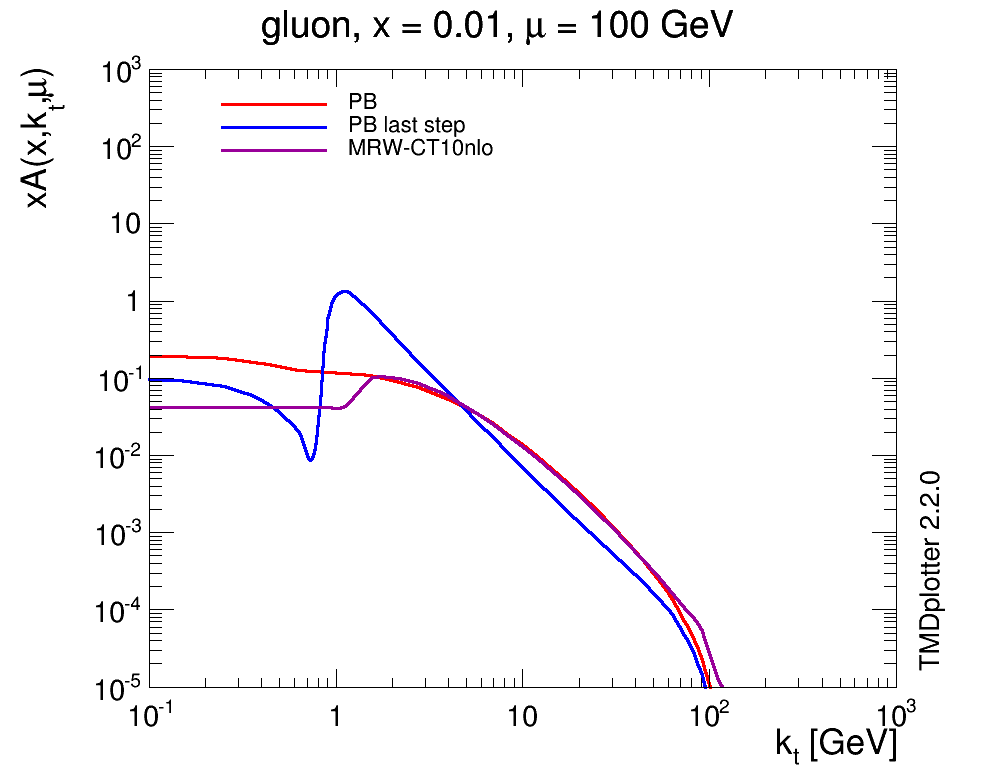}}
\end{minipage}
\begin{minipage}{0.49\linewidth}
\centerline{\includegraphics[width=0.99\linewidth]{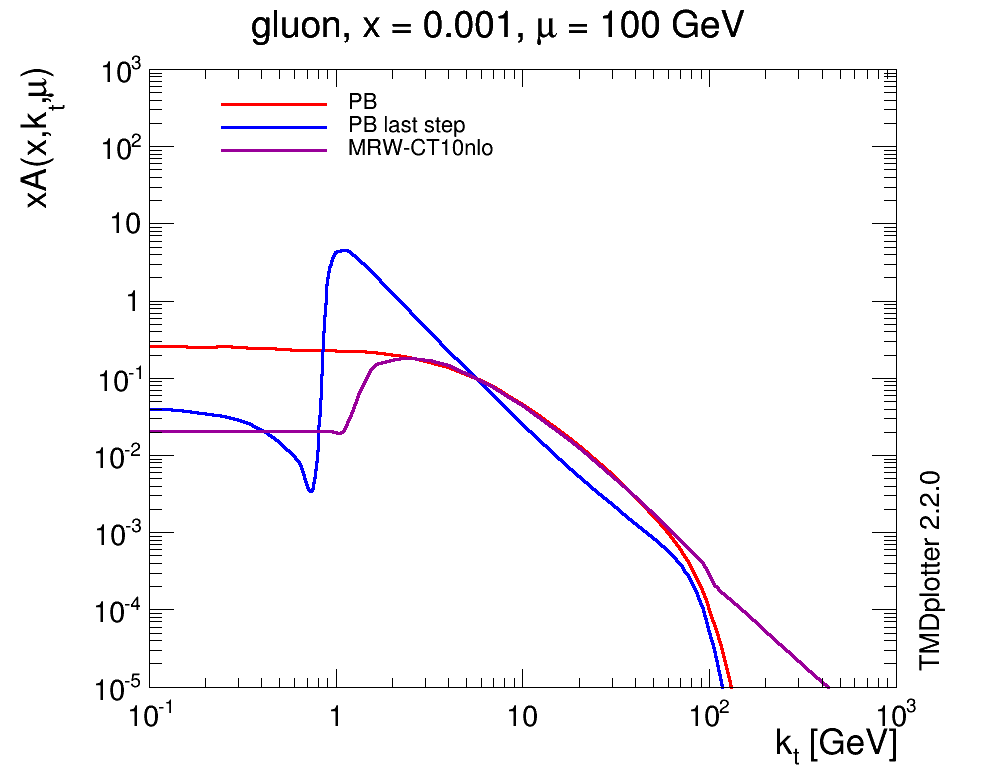}}
\end{minipage}
\begin{minipage}{0.49\linewidth}
\centerline{\includegraphics[width=0.99\linewidth]{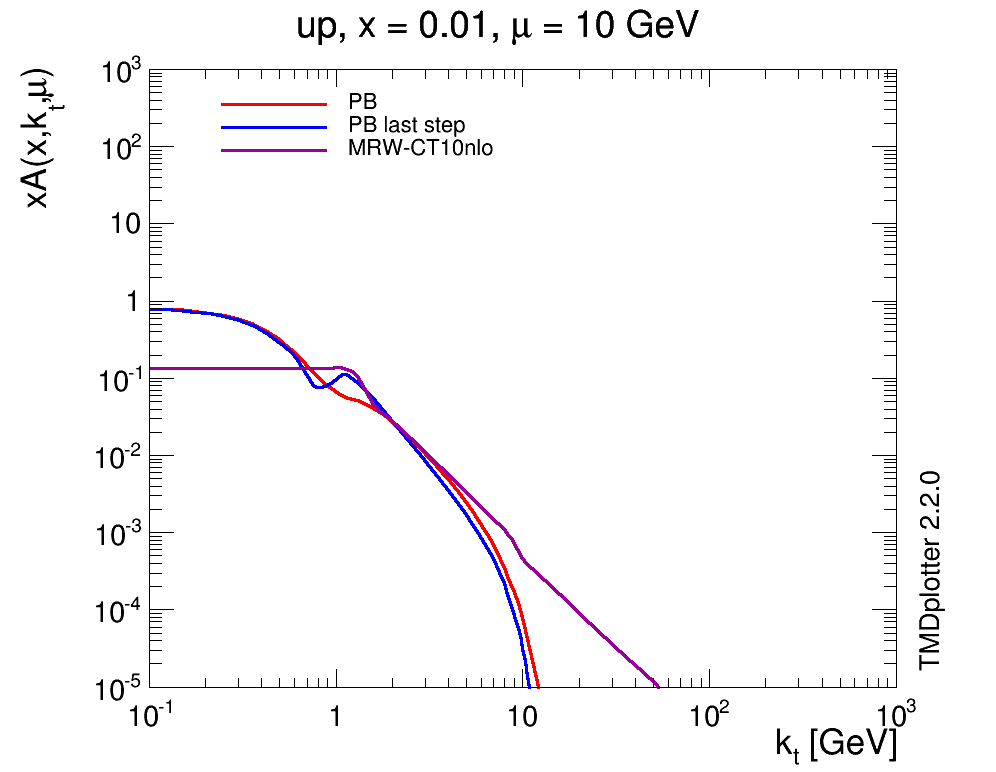}}
\end{minipage}
\hfill
\begin{minipage}{0.49\linewidth}
\centerline{\includegraphics[width=0.99\linewidth]{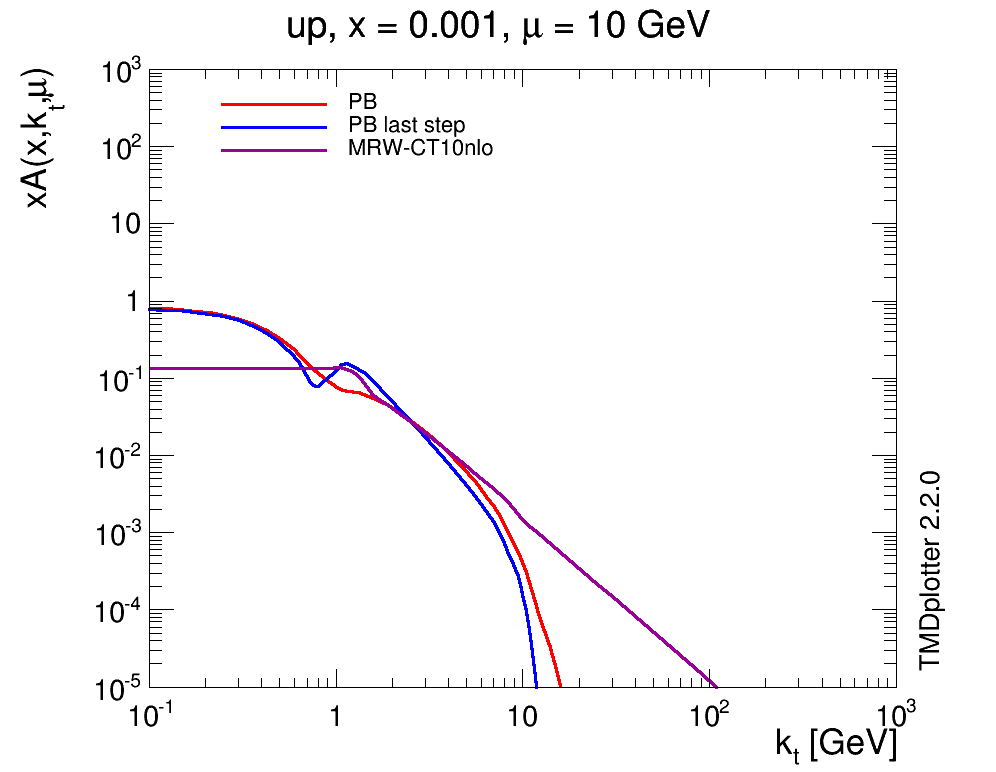}}
\end{minipage}
\hfill
\begin{minipage}{0.49\linewidth}
\centerline{\includegraphics[width=0.99\linewidth]{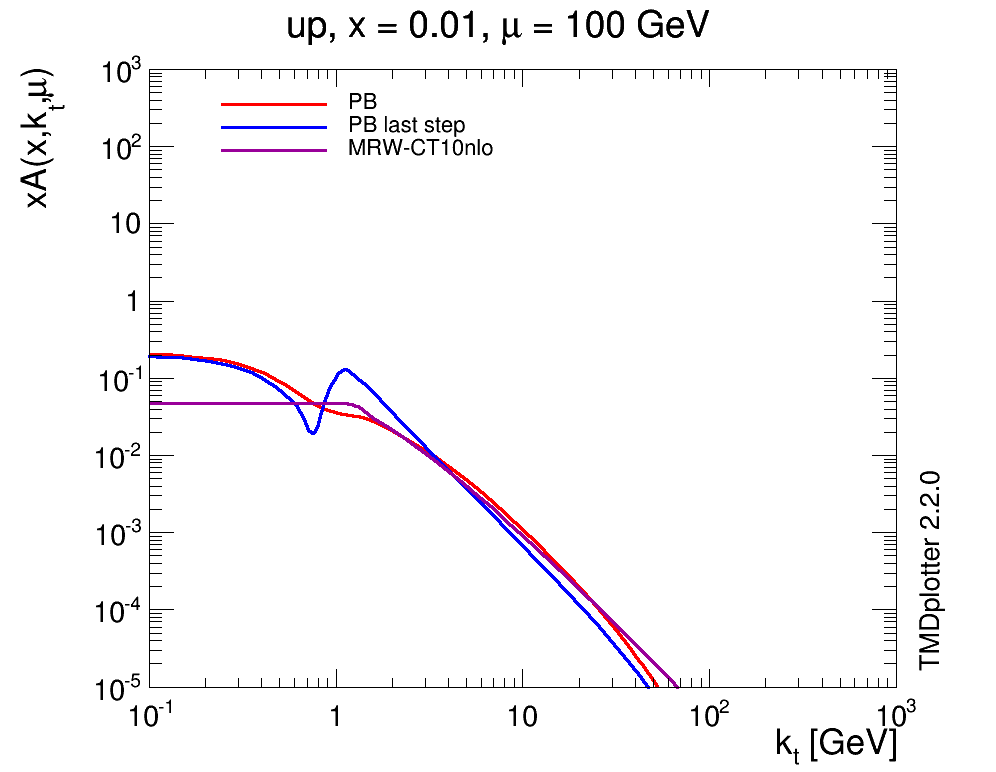}}
\end{minipage}
\begin{minipage}{0.49\linewidth}
\centerline{\includegraphics[width=0.99\linewidth]{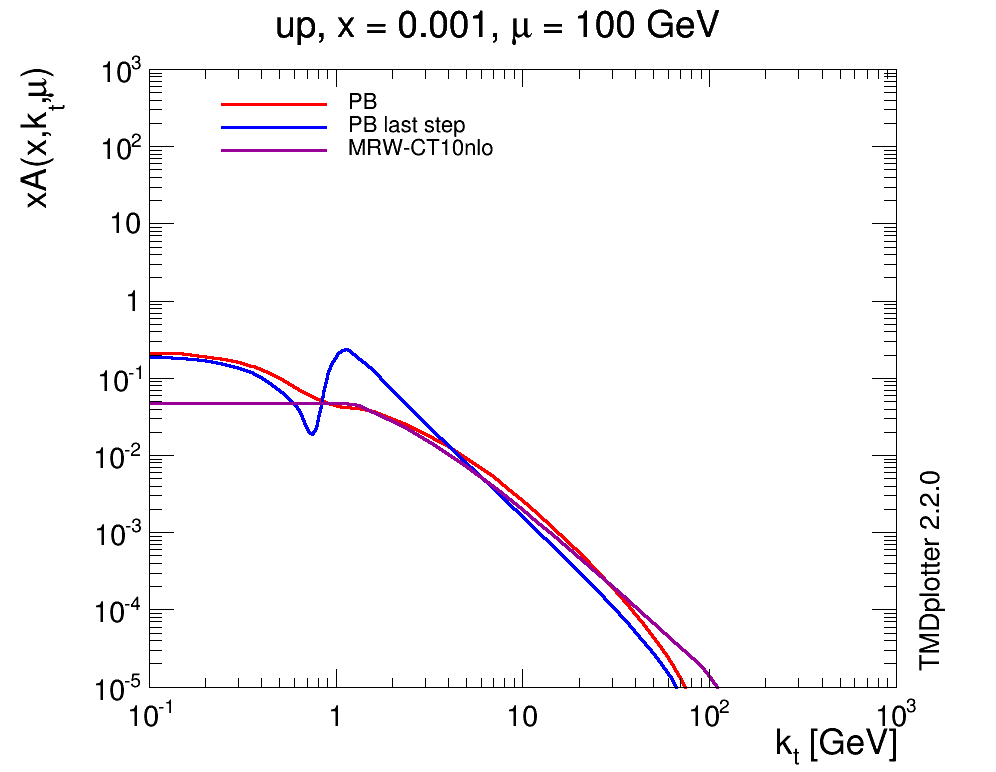}}
\end{minipage}
\hfill
\caption[]{TMDs from PB and KMRW as functions of transverse momentum for different parton species and 
different values  of longitudinal momentum fraction $x$ and evolution scale $\mu$.   }
\label{fig:KMRWvsPB}
\end{figure}
 
\begin{figure}
\begin{minipage}{0.49\linewidth}
\centerline{\includegraphics[width=0.99\linewidth]{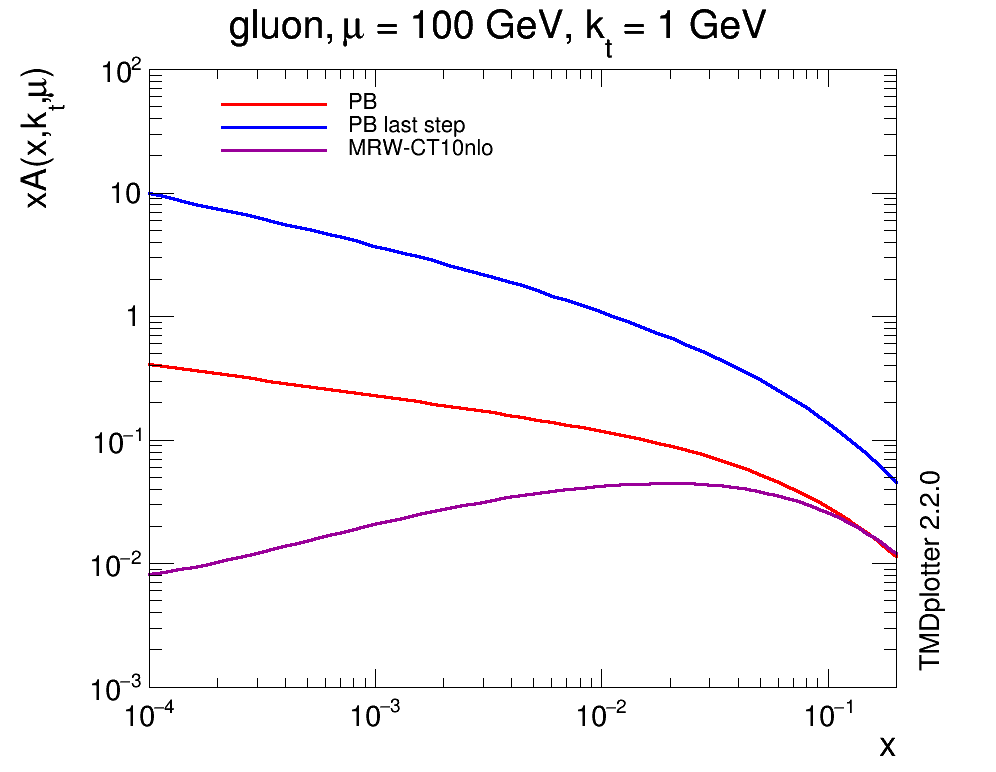}}
\end{minipage}
\hfill
\begin{minipage}{0.49\linewidth}
\centerline{\includegraphics[width=0.99\linewidth]{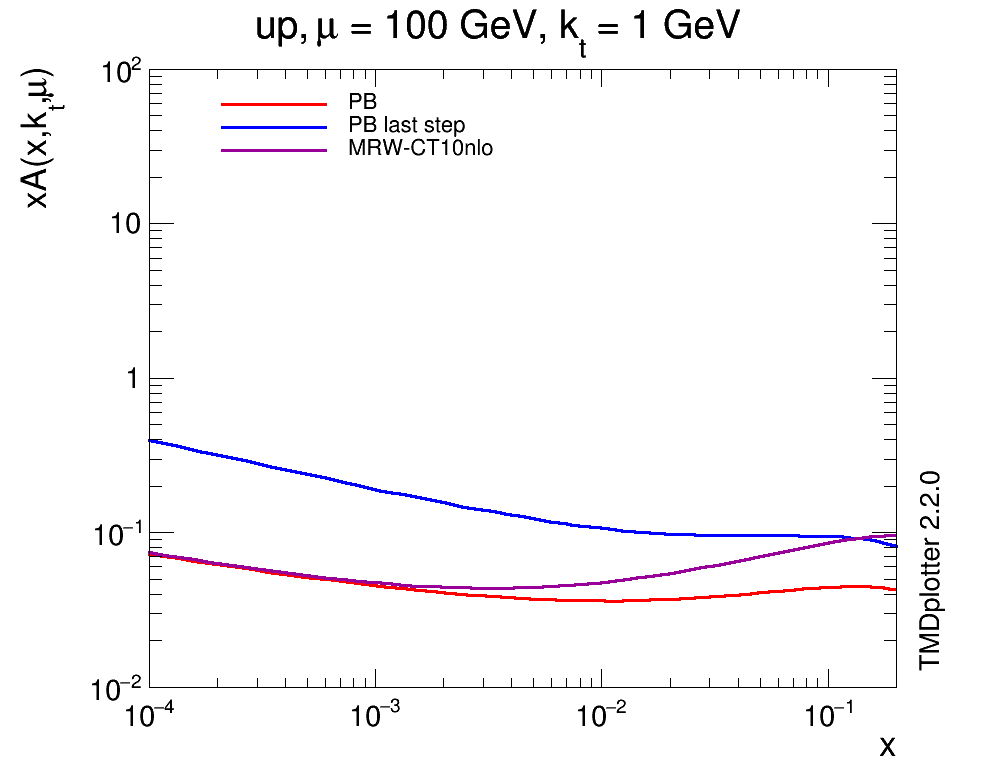}}
\end{minipage}
\hfill
\begin{minipage}{0.49\linewidth}
\centerline{\includegraphics[width=0.99\linewidth]{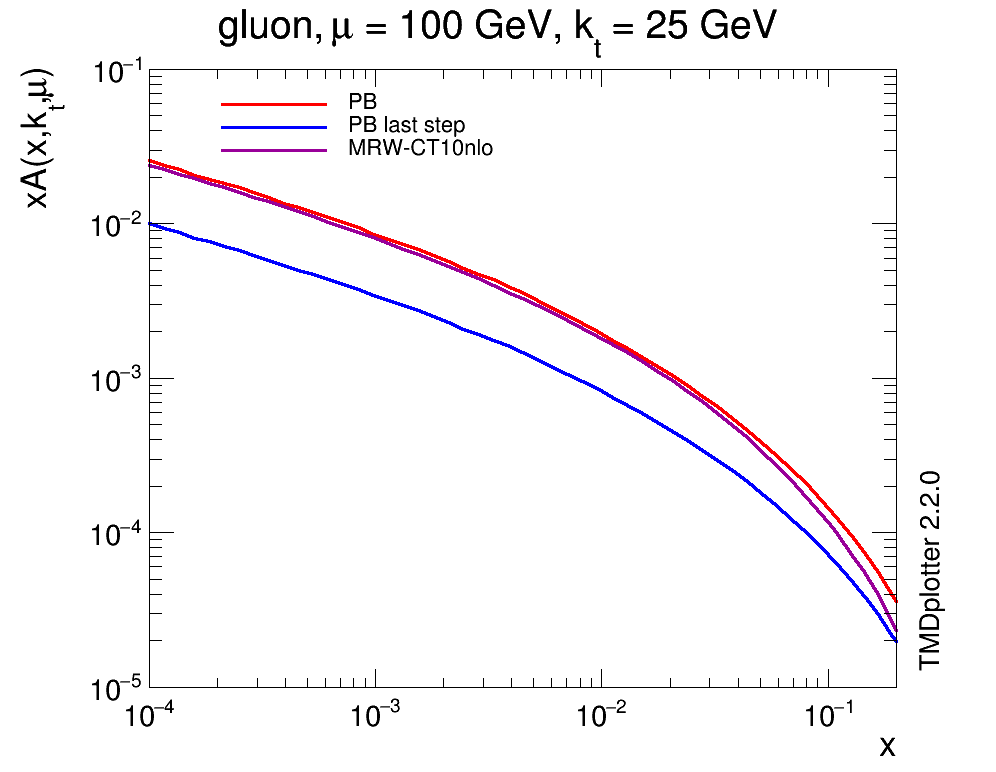}}
\end{minipage}
\hfill
\begin{minipage}{0.49\linewidth}
\centerline{\includegraphics[width=0.99\linewidth]{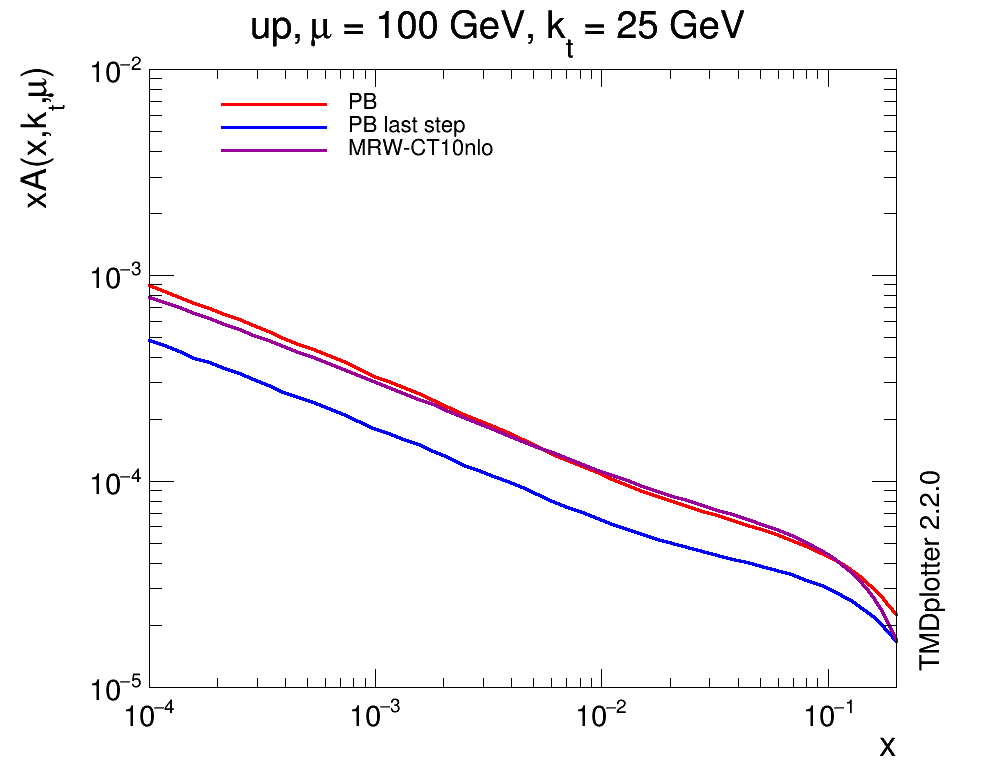}}
\end{minipage}
\hfill
\begin{minipage}{0.49\linewidth}
\centerline{\includegraphics[width=0.99\linewidth]{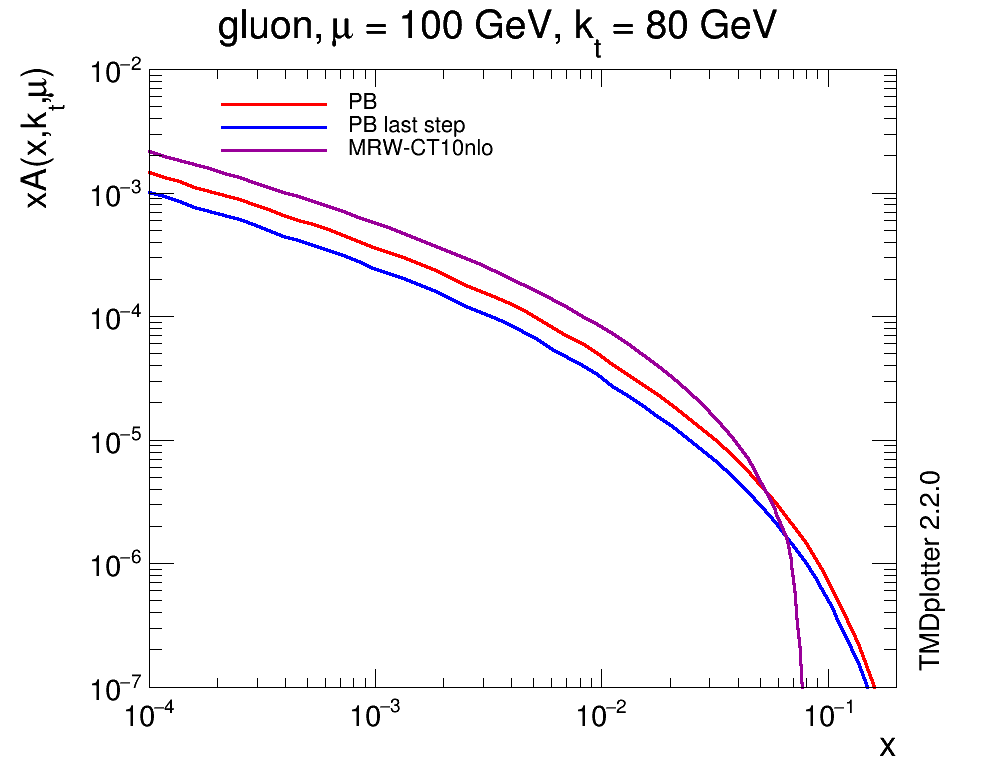}}
\end{minipage}
\hfill
\begin{minipage}{0.49\linewidth}
\centerline{\includegraphics[width=0.99\linewidth]{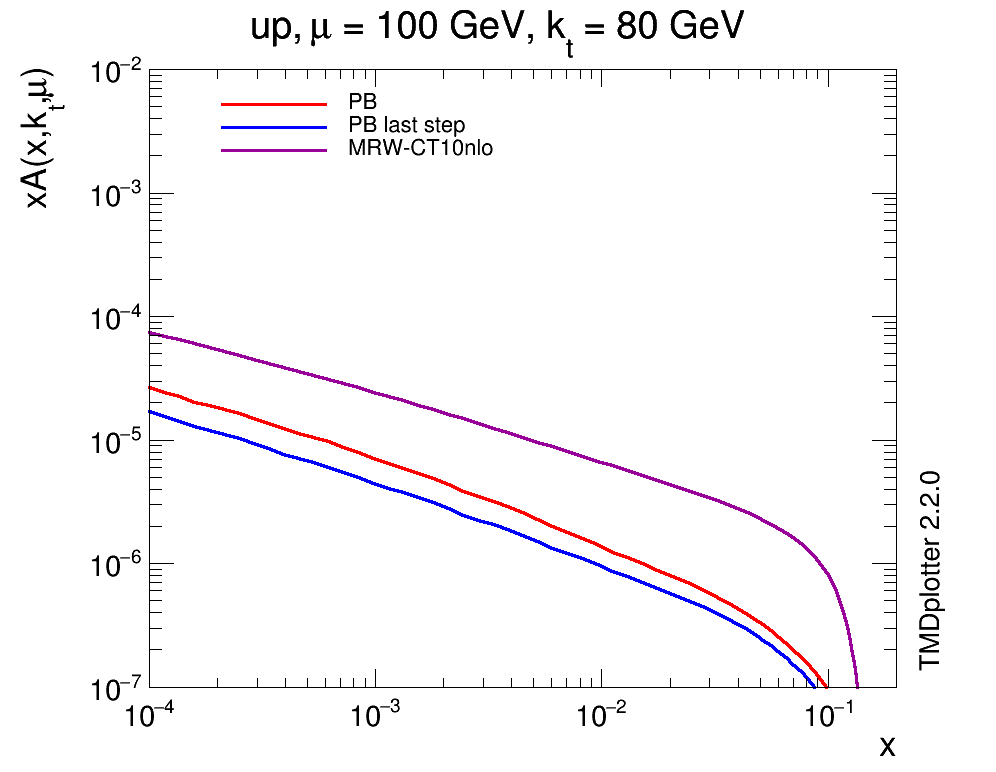}}
\end{minipage} 
 \caption[]{TMDs from PB and KMRW as  functions of $x$.   }
\label{fig:KMRWvsPB_vsx}
\end{figure}

\begin{figure}
\begin{minipage}{0.49\linewidth}
\centerline{\includegraphics[width=0.99\linewidth]{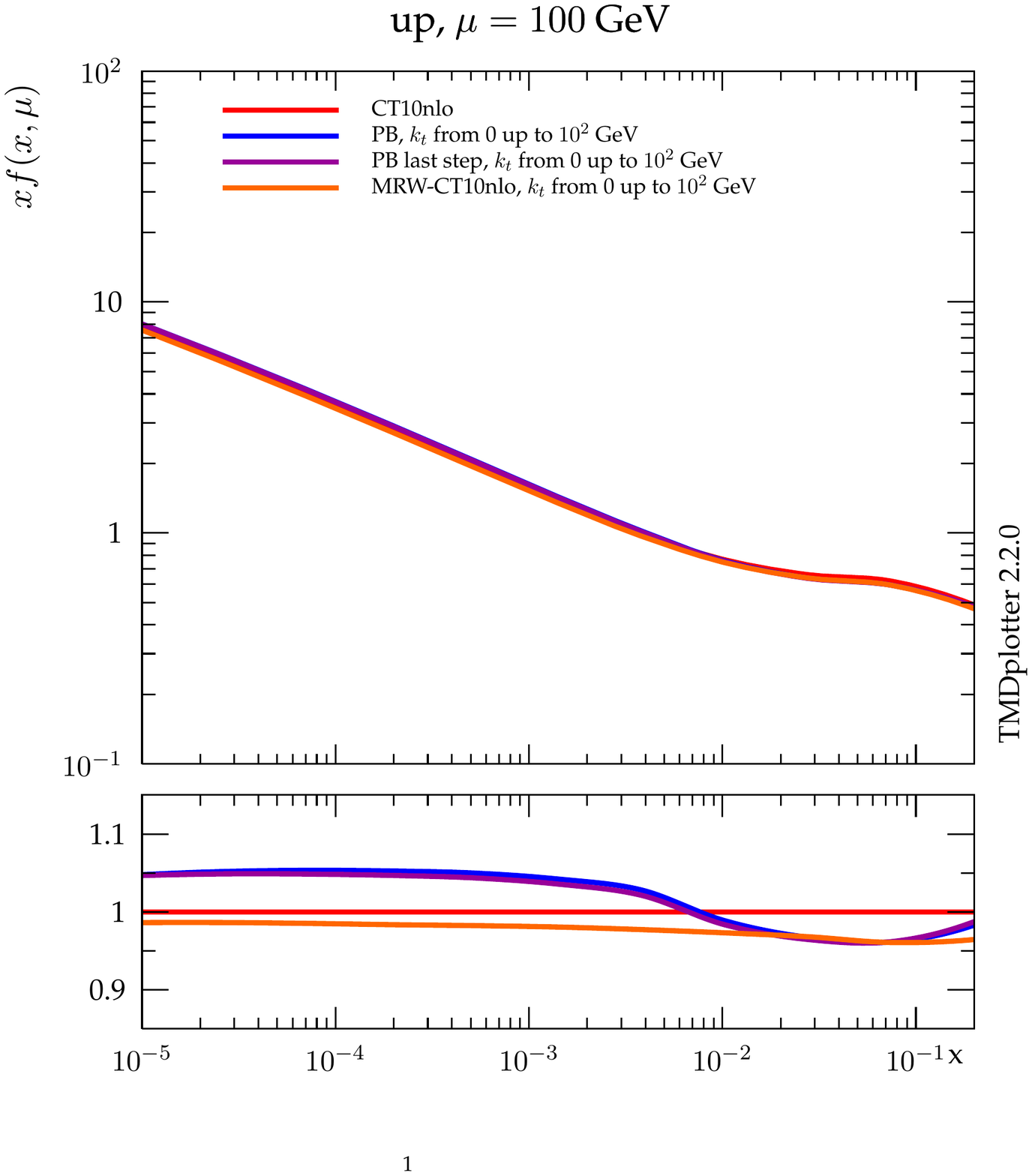}}
\end{minipage}
\hfill
\begin{minipage}{0.49\linewidth}
\centerline{\includegraphics[width=0.99\linewidth]{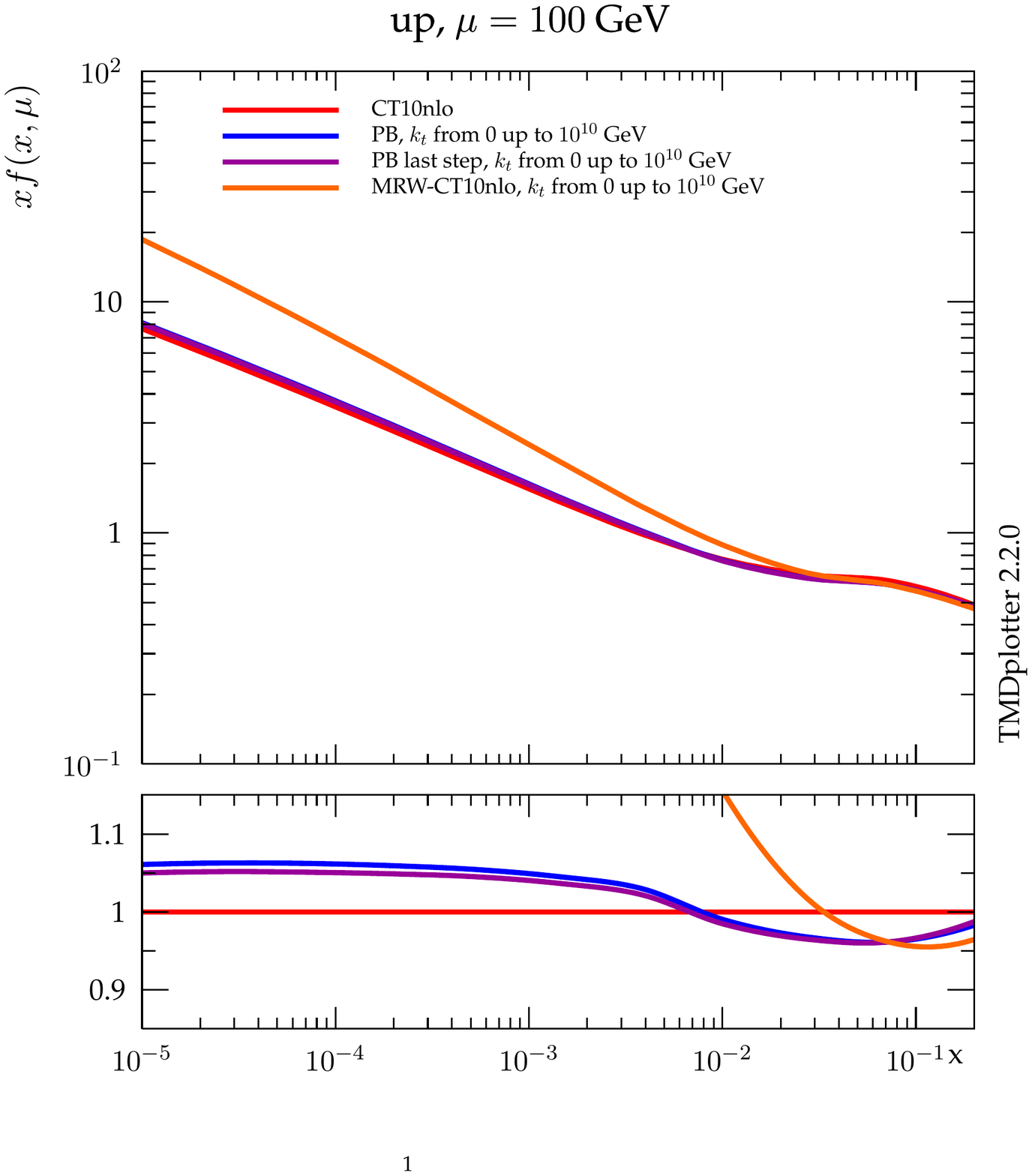}}
\end{minipage}
\hfill
 \caption[]{The results of integrating TMDs    over $k_{\bot} < \mu $ (left) and   over all $ k_{\bot}$ (right) as functions of $x$.}
\label{fig:KMRWvsPB_iTMDs}
\end{figure}

In Fig.~\ref{fig:KMRWvsPB} we show results for the $k_{\bot}$   dependence from PB, MRW-CT10nlo and PB-last-step calculations, at 
different  values of  $x$ and $\mu$.\footnote{The 
plots in Figs.~\ref{fig:KMRWvsPB}-\ref{fig:KMRWvsPB_iTMDs}   are obtained using the TMDplotter tool~\cite{Hautmann:2014kza,Connor:2016bmt}.} 
We may distinguish three regions of  low $k_{\bot}$, middle $k_{\bot}$ and high $k_{\bot}$,  characterized by distinct behaviors. 
Significant numerical  differences between PB and KMRW show up  especially in the extreme regions  $k_{\bot}  \ll  \mu $ and  $k_{\bot} \gg \mu $, while 
in an interval  of  middle  values  around $k_{\bot} \sim \mu$ the two predictions tend to become closer. 

In particular, we observe that at low $k_{\bot}$  the smearing of the intrinsic $k_{\bot}$ distribution due to evolution gives rise to different  behaviors in the 
single-emission and multiple-emission cases.   The kink at low $k_{\bot}$ in  MRW-CT10nlo  is a consequence of the single-emission picture, and is 
not present in the full PB case, where multiple branchings are responsible for generating the transverse momentum. 

We also observe that at high 
$k_{\bot}$ the MRW-CT10nlo distribution is far harder than PB and PB-last-step. This reflects the different pattern of radiative contributions in KMRW  from PB, 
illustrated in Sec.~4. As noted in~\cite{Bury:2017jxo}, the treatment of  the Sudakov form factor for $k_{\bot}^2>\mu^2$ influences the MRW-CT10nlo 
high $k_{\bot}$ tail. 

On the other hand, notice that 
 MRW-ct10nlo and PB are closer in the middle  range of  $k_{\bot}$ comparable  to the scale $\mu$. In this range, the differences between 
 KMRW and PB approaches  in the parton-density and  Sudakov-factor rescaling and phase space, discussed in the previous section, 
 compensate for  KMRW not taking into account all previous emissions compared to PB. The net effect is 
that the  behaviors of KMRW and PB   are  not too dissimilar for mid  $k_{\bot}$.
 
Results for the $x$  
  dependence from PB, MRW-CT10nlo and PB-last-step calculations are shown in Fig.~\ref{fig:KMRWvsPB_vsx},  illustrating  that the effects noted above 
persist over a broad range in $x$.

In Fig.~\ref{fig:KMRWvsPB_iTMDs} the results of integrating   MRW-CT10nlo, PB and PB-last-step TMDs 
over the transverse momentum $k_{\bot}$     
 at a given evolution scale $\mu$  are shown   as  functions of $x$.  Results are shown  for integrating 
 TMDs over $k_{\bot} < \mu $  (Fig.~\ref{fig:KMRWvsPB_iTMDs}(left))  and   over all $ k_{\bot}$ (Fig.~\ref{fig:KMRWvsPB_iTMDs}(right)). 
For comparison, we also plot   CT10nlo distributions at the same $\mu$. 
In the lower parts of the figure  the ratios of  integrated TMDs to  CT10nlo are plotted. 
As expected, we observe that 
none of the distributions integrate to CT10nlo, given that  the resolution scale $z_M$ is far from 1, and 
 the scale of the running coupling $\alpha_s$ is $q_{\bot}$  --- see remarks  below Eq.~(\ref{integratingAoverKt}).   
In the case of integrating over all $ k_{\bot}$ (Fig.~\ref{fig:KMRWvsPB_iTMDs}(right)) we  note that MRW-CT10nlo 
 gives rise to a much higher distribution than all other curves, implying that the MRW-CT10nlo  high-$k_{\bot}$ tail has a significant 
 impact at integrated level for most values of $x$. On the other hand, when integrating  over $k_{\bot} < \mu $  (Fig.~\ref{fig:KMRWvsPB_iTMDs}(left))  
 the deviation of MRW-CT10nlo from collinear CT10nlo is  smaller  than that of PB, which  is a further 
  manifestation  of the  differences between the KMRW and PB physical  pictures  illustrated in Sec.~4.

  \subsection{DY $Z$-boson $p_{\bot}$ spectrum}

$Z$-boson transverse momentum spectra in DY di-lepton production have been measured with high precision at 
the LHC~\cite{Aad:2015auj,Aad:2014xaa,Khachatryan:2016nbe,Chatrchyan:2011wt}. 
In the region of transverse momenta $p_{\bot}$ small compared to the di-lepton invariant mass, 
the spectrum is sensitive to Sudakov resummation. 
Reliable theoretical predictions require soft-gluon resummations and 
nonperturbative  contributions, which can be included by using the TMD theoretical framework. 
We here follow the treatment~\cite{Martinez:2018jxt} to apply PB TMDs to the DY $p_{\bot}$ spectrum. 
We obtain predictions for the $Z$-boson distribution based on PB TMDs that  include  
effects of dynamical soft-gluon resolution scale. We compare them with  KMRW  results.  
We perform a comparison of theoretical results  with the  LHC measurements~\cite{Aad:2015auj}. 
   
 Following~\cite{Martinez:2018jxt},  
as  we are interested mainly in the low $p_{\bot}$ region of the DY spectrum  we use on-shell LO matrix elements  (in the format of  Les Houches Event (LHE) 
file~\cite{Alwall:2006yp}) generated by  {\sc Pythia} Monte Carlo~\cite{Sjostrand:2007gs}. 
The transverse momentum of the initial state partons is calculated according to the TMDs and added to the 
event record in such a way that the mass of the produced DY pair is conserved, while the longitudinal momenta are changed accordingly. 
This procedure is common in standard parton shower approaches~\cite{Sjostrand:2014zea,Bengtsson:1986gz} and is implemented in 
the {\sc Cascade} package~\cite{Jung:2010si}.  Events in HEPMC~\cite{Dobbs:2001ck} format are  produced and  analyzed
with  Rivet~\cite{Buckley:2010ar}. 

\begin{figure}
\centerline{\includegraphics[width=9cm]{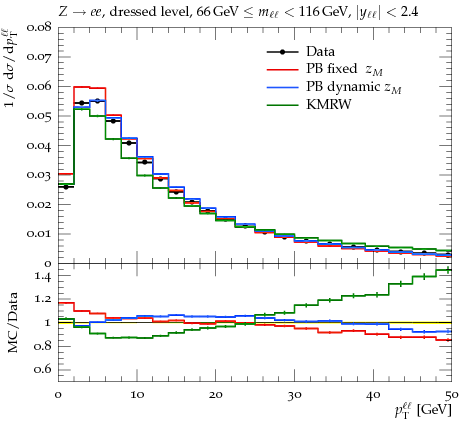}}
 \caption[]{ Predictions for the $Z$-boson $p_{\bot}$ spectrum obtained with PB  and MRW-CT10nlo TMDs compared to 
 the $8\;\textrm{TeV}$ ATLAS measurement~\protect\cite{Aad:2015auj}.}
\label{fig:Zpt}
\end{figure}

In Fig.~\ref{fig:Zpt}  predictions for the Z boson $p_{\bot}$ spectrum at  the LHC with $\sqrt{s}  = 8\;\textrm{TeV}$  
are shown using   MRW-CT10nlo and PB  TMDs,  and 
compared to the ATLAS measurements~\cite{Aad:2015auj}. For reference, we also plot PB results  for fixed (non-dynamical) resolution 
scale $z_M$  using the PB TMD Set-2 of Ref.~\cite{Martinez:2018jxt}. 
We see that the  MRW-CT10nlo calculation and PB calculation with dynamical $z_M$ give rise to different shapes in the $Z$-boson $p_{\bot}$ spectrum both 
in the region of low $p_{\bot}$  around the peak    and in the region of 
 high $p_{\bot}$ toward the upper end of the transverse momentum range shown.  There is an  interval of intermediate  $p_{\bot}$ in which they are less dissimilar. 
The agreement of the PB calculation with the measurements is good, while MRW-CT10nlo does not describe  the  high $p_{\bot}$ region, and the slope at 
low   $p_{\bot}$. 

Fig.~\ref{fig:Zpt}  also illustrates the effect of the soft-gluon dynamical  resolution scale by comparing PB predictions with fixed and dynamical $z_M$. 
We see that the slope of the  $p_{\bot}$ spectrum  is affected by dynamical  $z_M$  particularly in the low   $p_{\bot}$ region. 
The results indicate  that measurements of the $Z$ boson $p_{\bot}$ with high resolution in  the region  $p_\perp \ltap $ 5 - 10 GeV 
will  allow one to probe quantitatively effects of soft-gluon angular ordering and dynamical resolution scales. 
This will also be relevant to investigate effects from transverse momentum dependence in the branching probabilities 
(see e.g.~\cite{Hautmann:2012sh,Hautmann:2012pf}).  

We have limited ourselves to showing results for central values of the predictions, because TMD uncertainties in  the case of 
dynamical $z_M$ are not yet available.    The results obtained provide a strong motivation for 
  extending  the PB TMD fits and determination of TMD uncertainties \cite{Martinez:2018jxt}  
to include  dynamical  resolution scales.  We leave this to future work.

\section{Conclusions}
 
In this paper, using  the PB method~\cite{Hautmann:2017xtx,Hautmann:2017fcj} 
 for angular-ordered TMD evolution,  we have studied physical implications  of  the 
 dependence of the soft-gluon resolution parameter on the  branching scale.  
 Mapping the phase space of resolvable and non-resolvable emissions from 
  $(\mu^\prime , z)$ space to $(z , q_{\bot})$  space,   
 we have  written down the corresponding form of the evolution kernel. 
We have  established  the comparison of the PB formulation with 
other  existing formulations, notably the ones known as CMW~\cite{Marchesini:1987cf,Catani:1990rr}  
and KMRW~\cite{Kimber:1999xc,Kimber:2001sc,Watt:2003mx,Martin:2009ii}.  

On one hand, we find  that the PB formula coincides with CMW  at the level of integrated distributions.  CMW was 
originally developed by evaluating  splitting kernels at LO,  while we evaluate the kernels at NLO. On the other 
hand, we find significant differences of PB with respect to KMRW, which can be traced back to the fact that 
PB builds the initial-state transverse momentum from multiple emissions, while KMRW builds  it from single 
emission ---  the last step in the initial-state evolution cascade.   We  examine these  differences in detail 
both analytically and numerically.  We  
 find that the numerical effects are large in the extreme regions of low $k_\perp$ and high $k_\perp$, but 
small in the middle $k_\perp$ region. 

We apply the results to the evaluation of the DY $Z$-boson $p_\perp$ spectrum and comparison with 
LHC measurements. We compare PB versus KMRW, finding 
significantly different behaviors in the low-$p_\perp$ and high-$p_\perp$ regions. We   
 study the  sensitivity of the  $Z$-boson  spectrum  to
 effects of the soft-gluon resolution scale, and    
 observe  that these could be accessed by detailed measurements  of the 
  $Z$-boson transverse momentum with fine binning in the  region $p_\perp \ltap $ 5 - 10 GeV.

\vskip 0.8cm  

\noindent 
{\bf Acknowledgments.} We thank M.~Bury for helpful correspondence on the  MRW-CT10nlo implementation, and  
 K.~Golec-Biernat, H.~Jung and S.~Pl{\" a}tzer for useful discussions.   We  gratefully acknowledge the  hospitality and support of 
  the Erwin Schr{\" o}dinger Institute at the University of  Vienna and 
   Nuclear Physics Institute of the Polish Academy of Sciences in Krakow 
   while part of this work was being done.   
 
 \bibliographystyle{mybibstyle} 

\renewcommand{\bibname}{\hsppp\textcolor{gray75}{|}\hsppp Bibliography}
 \newpage
    {

\bibliography{PracaDok17aug}
    \cleardoublepage

    }

\end{document}